# A microelectromechanically controlled cavity optomechanical sensing system

**Houxun Miao[1,2], Kartik Srinivasan[1] and Vladimir Aksyuk[1,3]**
[1] Center for Nanoscale Science and Technology, National Institute of Standards and Technology, 100 Bureau Dr, Gaithersburg, MD 20899, USA
[2] Maryland Nanocenter, University of Maryland, College Park, MD 20742, USA
E-mail: vladimir.aksyuk@nist.gov

**Abstract.** Microelectromechanical systems (MEMS) have been applied to many measurement problems in physics, chemistry, biology and medicine. In parallel, cavity optomechanical systems have achieved quantum-limited displacement sensitivity and ground state cooling of nanoscale objects. By integrating a novel cavity optomechanical structure into an actuated MEMS sensing platform, we demonstrate a system with  high quality-factor interferometric readout, electrical tuning of the optomechanical coupling by two orders of magnitude, and a mechanical transfer function adjustable via feedback. The platform separates optical and mechanical components, allowing flexible customization for specific scientific and commercial applications. We achieve displacement sensitivity of 4.6 fm/Hz$^{1/2}$ and force sensitivity of 53 aN/Hz$^{1/2}$ with only 250 nW optical power launched into the sensor. Cold-damping feedback is used to reduce the thermal mechanical vibration of the sensor by 3 orders of magnitude and to broaden the  sensor bandwidth by approximately the same factor, to above twice the fundamental frequency of $\approx$ 40 kHz. The readout sensitivity approaching the standard quantum limit is combined with MEMS actuation in a fully integrated, compact, low power, stable system compatible with Si batch fabrication and electronics integration.

[3] Author to whom any correspondence should be addressed.

# 1. Introduction

Microelectromechanical system (MEMS) sensors operate by transducing physical quantities such as acceleration, force, mass or magnetization, into measurable mechanical motion that is suitably detected [1-6]. Their small size, low power consumption and low-cost batch-production lead to a large range of commercial applications in pressure, acoustic, inertial sensing, and atomic force microscopy (AFM) [7]. MEMS enable exquisite and unique measurements at the forefront of science: detecting individual flux lines in superconductors [8]; sensing small numbers of nuclear spins in magnetic resonance force microscopy [9]; measuring mechanical torque from electron spin flips [10]; and quantifying the Casimir force [11-13] from zero point fluctuations of the electromagnetic vacuum. In all these examples, precise and accurate readout of the mechanical motion of the microscopic sensor plays a critical role.

While a variety of readout techniques has been developed and used, the ideal combination of precision, bandwidth, low power dissipation, small sensing area and on-chip integration remains elusive, imposing significant performance constraints. While care is taken to reduce the mechanical dissipation of the sensor and the corresponding thermal Langevin force noise of the first, mechanical transduction stage, the noise of the second, readout stage often dominates, particularly away from the narrow frequency region of high responsivity near the mechanical resonance peak. Optical readout is attractive because it can have a small sensing area on the order of the wavelength squared, and introduce no extra thermal noise, being shot noise limited. Its enormous bandwidth is constrained, in principle, only by the optical frequency. Cavity-enhanced interferometry [14-16] trades off that bandwidth for the resonant enhancement of sensitivity by the optical quality factor ($Q$) which can be as high as $10^6$ or more. This results in a remarkably low readout noise even at very small optical power. This principle has been applied to a laser interferometer gravitational-wave observatory for gravity wave detection [17], to micro systems for quantum-limited displacement sensitivity [18-21], optical force and radiation pressure studies [22-28] and fast AFM [29]. However, it remains difficult to couple soft, sensitive mechanical transducers with stiffness and resonance frequency on the order of 10 mN/m and 10 kHz with high finesse optical cavities, and, particularly, to fully integrate these components into highly stable and batch-produced single-chip sensors.

Here we fully integrate a novel high $Q$ cavity optomechanical position sensor into an electrostatically actuated MEMS device [30-32] and demonstrate high-sensitivity position readout in a single-chip, fiber-pigtailed, low-power, compact and stable microsystem compatible with batch fabrication at low cost. Conceptually similar to other MEMS sensors [9-13], the device includes a movable surface where samples can be attached or unknown forces or torques applied, while the mechanical response is now detected optically with precision and bandwidth improved by orders of magnitude compared to a more conventional electrostatic or piezoresistive readout. Importantly, the ability to actuate the device electrostatically is retained. The combination of extremely low noise readout and MEMS electrostatic tuning results in several new system capabilities. The optical cavity resonance is tunable through 5.54 nm, useful for adjusting the device to operate with a fixed wavelength optical laser source, such as a compact and relatively inexpensive stabilized laser diode. The optomechanical coupling and thus the readout sensitivity is tunable by over 2 orders of magnitude for optimizing the sensor gain and dynamic range.

For the first time in a fully integrated MEMS nanophotonic device, we apply electronic feedback and demonstrate cold damping of the mechanical degree of freedom [20, 33-36] by more than 3 orders of magnitude. While not reducing the input-referred force noise, i.e. the Langivin force acting on the sensor, the near critical damping stabilizes the sensor position and also flattens the frequency-dependent sensor gain, allowing us to use the sensor effectively over a very broad frequency range without severe dynamic range constraints. To demonstrate this, we measure the force noise acting on the sensor for frequencies up to 100 kHz, more than 2.5 times the fundamental resonance frequency of 38.76 kHz. Even with the low mechanical responsivity in this high frequency range the input referred readout noise remains significantly below the thermal mechanical force noise being measured..

The ability to dampen the mechanical noise can strongly reduce the backaction of the sensor onto a system being measured. Moreover, our noise level is ≈ 2.3 times the standard quantum limit (SQL) [19, 37] for our mechanical system, approaching the fundamental readout limits. With future parameter improvements, cooling the sensor to the quantum mechanical ground state while maintaining the high readout bandwidth [38] (i.e. in the 'bad' cavity limit) may be achieved.

## 2. Results

Figure 1 shows the scanning electron micrographs (SEM) and the illustrated cross sections of the sensing system. The detailed fabrication process is described in Appendix A. The 15 μm diameter, 240 nm thick stationary silicon microdisk optical resonator is used to detect the out of plane mechanical motion of a 200 nm thick, 19 μm outer diameter silicon nitride ring suspended a variable distance $Z$ above. The resonant optical modes of the microdisk are excited and measured by an integrated single mode on-chip waveguide, which has been fiber-pigtailed for robust and low-loss coupling of light into and out of the device. The microdisk is fixed to the substrate via a silicon nitride anchor, while the ring is attached to a MEMS actuator. The actuator is a silicon cantilever fixed to the substrate via silicon nitride anchors on one side and connected to the ring on the other side. The cantilever consists of two electrically separate parts that are mechanically joined by a dielectric silicon nitride bridge. When a voltage is applied between the first part of the cantilever and the substrate (1 μm below), the attractive electrostatic force bends it towards the substrate. The second part of the cantilever, maintained at the ground potential, serves as a mechanical lever arm to achieve a larger range of motion at the ring location. The electrostatic tuning of $Z$ as a function of the applied DC voltage is shown in Fig. 2a, measured with a white light interferometer.

Having the movable parts mechanically separated from the optical readout components (disk and waveguide) allows us to independently choose the key parameters of both the optical sensing and mechanical transduction. The silicon disk resonator operates with an intrinsic quality factor of nearly $10^6$, approaching the highest values achieved in silicon nanophotonic resonators [39, 40], and contributing to the optical sensor's high sensitivity to the ring's motion. Through adjustment of the waveguide-cavity separation, the loaded quality factor of the system can be tuned, determining the optical bandwidth and ensuring that the out-coupled optical signal into the waveguide is sufficiently strong to allow near shot-noise-limited detection. The mechanics is similarly largely unconstrained by the optical readout. The parameter space of mechanical frequency, stiffness, and ring displacement range can be adjusted depending on the specific sensing application at hand. For example, in the current work, wide readout tunability afforded by the relatively large ring displacement range was the dominant consideration. For other applications, working at a pre-determined disk-ring gap that is chosen based on the desired readout gain and dynamic range would allow using stiffer cantilevers with higher mechanical frequencies. The sensor platform is also compatible with torsional, parallel plate or elastic membrane mechanical elements and integrated magnetic, thermal or piezoelectric actuators.

As illustrated in Fig. 1c, in a thin microdisk resonator a significant portion of the optical mode energy is located in the evanescent tails below and above the disk. When the gap Z between the nitride ring and the disk is reduced, the mode shape and frequency are strongly modified. Similar to a change in the position of one of the mirrors in a Fabry Perot cavity, a small change in the position of the nitride ring shifts the optical resonance frequency linearly with mechanical displacement. The shift can be measured by sensing the modulation of continuous wave (c.w.) light tuned to near cavity resonance, accomplished either by simple amplitude measurement on the shoulder of the resonance or a suitable phase sensitive measurement, such as polarization spectroscopy [41] or Pound-Drever-Hall technique [42].

Figure 2b shows the transmission spectrum of one of the optical modes that is used for subsequent experiments throughout the paper; the loaded optical $Q = 3.8 \times 10^5$ is estimated from the linewidth of the mode as shown in the figure. This mode is of transverse magnetic (TM) polarization, so that the electric field is predominantly oriented normal to the plane of the microdisk. Figure 2c shows the resonant wavelength as a function of the applied voltage, measured by varying the laser wavelength while the cavity is locked onto the laser via feedback (see Fig. 3 and discussion below). An active tuning of the cavity resonance over the range of 5.54 nm ±0.02 nm is achieved. The uncertainty is the laser wavelength tuning resolution in this measurement. The system is highly repeatable and stable. The tuning curve is reproducible within the wavelength accuracy of 0.02 nm of the laser after two weeks. This corresponds to an upper bound on the average slow drift of the gap of 0.1 nm per day. The sensor readout gain is proportional to the optomechanical coupling $g_{OM}=d\omega_c/dZ$, where $\omega_c$ is the optical resonance frequency. It is calculated from $\omega_c$ and $Z$ measurements in Fig. 2a and 2c, and shown in Fig. 2d. The optomechanical coupling is increased by a factor of more than 200 (from $g_{OM}/2\pi$ = 65 MHz/nm to $g_{OM}/2\pi$ = 13.4 GHz/nm) when the applied voltage is increased from 2.5 V to 8 V, thus broadly adjusting the gain of the readout. The tuning can also be used to match the cavity to a pre-set working wavelength, e.g. so that an inexpensive fixed wavelength diode laser source can be used for sensing.

We use the cavity optomechanical sensor to study the dynamics of the fundamental mechanical cantilever bending mode (inset in Fig. 1a). The device is placed in a vacuum chamber, where a pressure of ≈ 0.3 Pa is maintained to reduce the mechanical loss due to air damping. In an optical transmission measurement (Fig. 3), a c.w. wave from a tunable laser is launched into the device, with an estimated optical power of ≈250 nW in the waveguide right before the microdisk, and ≈40 nW out-coupled to a photodetector with a gain of ≈1.9x10$^6$ V/W and the 3dB bandwidth of 200 kHz. The wavelength is tuned to a pre-specified transmitted power set point as shown in Fig. 2b. A small motion of the cantilever results in a linear modulation of the transmitted optical power via the optomechanical coupling. The resulting photodetector voltage proportional to the mechanical displacement is recorded with an electrical signal analyzer. The calibration to obtain the absolute value of the mechanical displacement at the center of the ring is described in Appendix B.

We have measured two characteristics of our device under various conditions: the mechanical displacement noise spectral density and the sensor transfer function, shown in Fig. 4a and 4b, respectively. The transfer function quantifies the linear response of the MEMS sensor to an external applied force at various frequencies. It is obtained by applying an additional small sinusoidal voltage to the actuator to mimic an external force and recording the resulting displacement as shown in Fig. 3. It is calibrated in terms of both the applied force and the displacement at the center of the ring, where the actuator mechanical stiffness in the $Z$ direction is estimated from the finite element modeling (FEM) to be $k_0$=0.038 N/m and is dominated by the fundamental mechanical mode.

The blue curves on Fig. 4a and 4b are taken under a fixed actuator bias voltage of 3V, with the inset in Fig. 4a showing the high resolution data near the resonance frequency. The fundamental mechanical mode has a frequency of 38.52 kHz, which agrees well with 38.7 kHz from FEM. The inset in Fig. 4b shows the normalized transfer function under the same bias, measured with a sufficiently low optical power (an additional 26 dB attenuation compared to Fig. 4a) to extract the intrinsic mechanical $Q$. At this power level, radiation pressure induced optomechanical spring, as well as excitation or damping are negligible, so that the values measured for both the blue and red detuning of the laser from the center of the cavity resonance agree with each other. The mechanical frequency and the intrinsic $Q$ are estimated to be $\approx$ 38.76 kHz and $\approx$ 1400, respectively.

The black curve in Fig. 4a shows the noise background of the measurement, obtained by detuning the laser from the optical resonance. Very similar values are obtained by completely turning the laser off, indicating that the dominant noise source in our experiment is the detector dark noise. At the moderate optomechanical coupling achieved at 3V DC bias the sensor is dominated by the mechanical noise at *all* frequencies below about 50 kHz and by the readout noise at higher frequencies. As shown below, with higher optomechanical gain the input referred readout noise is further reduced and the sensor becomes limited by the mechanical noise over an even larger bandwidth of 0 to 100 kHz.

While the high mechanical $Q$ is desirable, corresponding to lower losses and lower thermal mechanical noise, the resulting highly non-uniform transfer function severely limits the sensor dynamic range if broadband application is considered, e.g. when small forces off resonance have to be measured in the presence of on-resonance forces. The on-resonance forces and thermal noise are mechanically amplified and can exceed the linear dynamic range of the readout set in our measurement by the linear portion of the shoulder of the optical resonance curve. In fact, as the optomechanical coupling is increased by increasing the bias voltage, e.g. $g_{OM}/2\pi$=0.24 GHz/nm at 5 V, our estimated thermal root mean square (rms) noise at room temperature of $(k_bT/k_0)^{1/2} \approx$ 0.47 nm will result in a cavity shift comparable to the cavity linewidth of 4.1 pm.

We overcome this limitation by introducing a cold damping feedback loop to make the system transfer function more uniform across the frequency spectrum and drastically reduce the cantilever displacement noise. Cold damping of the mechanical mode is realized by applying to the integrated actuator an electrical feedback signal derived from the optical readout signal. The feedback loop illustrated in Fig. 3 consists of two components implemented with two amplifier/filter chains. The top chain responds at frequencies from DC to 100 Hz and provides a negative proportional-integral feedback to the actuator to lock the cavity onto the laser and follow the laser wavelength when it drifts or is tuned intentionally. The set point is ≈ 43 % of the off resonance power on the red side of the optical mode (as indicated in Fig. 2b), the same throughout the paper. The bottom chain in Fig. 3 is used for cold damping. It provides a positive proportional feedback at frequencies above 300 Hz. Through the combination of two low pass filters (the detector at 200 kHz and subsequently another low pass filter at 300 kHz) the signal in this arm acquires a time lag that results in a sufficient out-of-phase component to provide the damping signal to the actuator.

The red curves in Fig. 4a and 4b show the noise spectrum and the corresponding transfer function at the same bias of 3V with feedback cooling. The green curve in Fig. 4b illustrates the transfer function with a higher damping gain setting, where the mechanical $Q$ is approaching 1. Considering the measured intrinsic mechanical $Q$ of ≈1400, the mechanical mode is damped by more than 3 orders of magnitude, decreasing the noise-driven rms sensor position variation by the same factor.

Stabilizing the sensor with feedback allows working at increased DC bias, reaching higher optomechanical coupling for higher readout gain and lower readout noise, which in turn also ensures negligible excess noise is injected back into the system by the feedback. Figure 5 shows the displacement noise spectra and the measurement noise backgrounds at increasing DC bias voltages and readout gains. The displacement sensitivity increases by more than a factor of 15 when the bias voltage is increased from 4.5 V to 6.75 V, and is limited by the photo detector noise to (4.6 x $10^{-15}$ ± 0.6 x $10^{-15}$) m/Hz$^{1/2}$ (average value across the spectrum; errors represent one standard deviation throughout the rest of the paper (Appendix B)) at a bias voltage lever of 6.75 V. It should be noted that this displacement sensitivity is only 2.3 times larger than the standard quantum limit ($S_x^{SQL}=\hbar Q/k_0$) for our cantilever, and one can expect a further increase of the measurement sensitivity by using a photo detector with lower noise, and bringing the nitride ring even closer to the microdisk. As evident from the figures, at high voltage, the displacement signal is well above the detector noise over the whole DC to 100 kHz frequency range, which makes the system suitable for dynamic processes measurements up to the frequency in excess of 2.5 times the fundamental mechanical frequency of the device.

The degree of damping can be adjusted independently of the bias voltage, as illustrated by the two sets of curves in Fig. 5a and 5b. In addition to damping, this simple feedback choice also results in the reduction of the closed-loop mechanical stiffness and corresponding decrease of the resonance frequency. While beyond the scope of this work, employing a more sophisticated feedback scheme with a high bandwidth (>> 100 kHz) proportional-integral-derivative (PID) controller, or another general linear controller, may enable one to engineer flat or other desired transfer functions in the future. The increased displacement at low frequencies is caused by 1/f excess force noise in our devices.

Using the measured transfer functions corresponding to the displacement noise data in Fig. 5b, we have obtained the spectra of the force noise at the input of our sensor, shown in Fig. 6. (See Appendix B for detailed calculations). The force noise spectra consist of a 1/f noise dominant at lower frequencies and flat, white noise appearing at high frequencies. Based on the linear dependence on the actuator DC bias voltage, we attribute the 1/f force noise to 1/f electrical noise on the cantilever, probably due to poor electrical contacts between metallic probes and doped silicon electrical pads of the device. The white noise component agrees well with the expected intrinsic thermal Langevin force noise (dashed curve) of $1.3 \times 10^{-15}$ N/Hz$^{1/2}$ calculated using the fluctuation dissipation theorem with the mechanical loss based on a measured intrinsic $Q$ of $\approx 1400$. The dotted line in Fig. 6 corresponds to the force sensitivity limit imposed by the optical readout noise. The photo detector limited force sensitivity is estimated to be $(5.3 \times 10^{-17} \pm 0.2 \times 10^{-17})$ N/Hz$^{1/2}$ at $\approx 25$ kHz (resonant frequency with damping feedback) with a 6.75 V DC bias.

## 3. Summary and Discussion:

We have developed a novel integrated MEMS sensing platform enabled by cavity optomechanics. We demonstrate a displacement readout sensitivity of $(4.6 \times 10^{-15} \pm 0.6 \times 10^{-15})$ m/Hz$^{1/2}$ and a force readout sensitivity of $(5.3 \times 10^{-17} \pm 0.2 \times 10^{-17})$ N/Hz$^{1/2}$ with the optical power of only 250 nW. The sensitive low-noise optomechanical readout in combination with electrostatic actuation is used to cold-dampen the mechanical mode by more than 3 orders of magnitude, achieving a corresponding suppression of the displacement noise, flattening of the overall sensor transfer function and expanding the sensor bandwidth to 100 kHz, exceeding the initial mechanical resonant bandwidth of f/Q = 28 Hz by a factor of ≈ 3500 and the fundamental mechanical resonance frequency by a factor of ≈ 2.5. Because of the dramatically improved readout sensitivity, the force sensitivity of the system is set by the thermal mechanical noise limit even at these high frequencies. This new regime in MEMS sensing is demonstrated in a compact, fiber-pigtailed, low-power, stable microsystem compatible with silicon batch fabrication at low cost and CMOS integration. The flexibility to independently choose optical and mechanical parameters of the described platform allows the system to be tailored for various sensing tasks.

In the future, at least one order of magnitude improvement in the thermal force noise can be expected from optimizing the processing to lower mechanical dissipation, reaching mechanical Q factors of $10^5$ [43] reported for similar types of soft, sensitive, low frequency Si cantilever devices. Our sensing and integration approach will be able to take full advantage of such thermal noise improvement for fast, broadband force measurements that are even more sensitive.

Our current position readout noise is a factor of 2.3 ± 0.3 above the standard quantum limit at the resonance frequency for our mechanical system and can be further improved by increasing the detected optical power, utilizing a lower noise detector, and further increasing the optomechanical coupling. With cryogenic cooling and increased mechanical Q for longer decoherence time, this approach may enable SQL-level readout with the bandwidth wider than the inverse decoherence time. In this regime quantum mechanical behavior of the system is observable. In contrast to optical cooling where a 'good' cavity (cavity linewidth narrower than the mechanical resonance frequency) is required, cold damping in principle allows reaching the quantum mechanical regime in the 'bad cavity limit', with large optical cavity resonance linewidth [38], thus maintaining the readout bandwidth in excess of the mechanical frequency. This not only relaxes the very stringent requirements on the optical cavity quality factor for the mechanically sensitive, low-stiffness, low frequency mechanical systems, but could open a new regime in quantum limited mechanical sensing with large readout bandwidth.

**Appendix A**

In this appendix, we describe the device fabrication process. We start with a silicon-on-insulator (SOI) wafer with 240 nm top silicon layer and 1 μm buried oxide (BOX) layer. The Si layer is patterned via electron beam lithography and reactive ion etched (RIE) with a $SF_6+C_4F_8$ recipe down to the BOX layer to produce high Q silicon microdisks, access waveguides for coupling light to/from the microdisk, and the actuators. The width of the waveguide is 500 nm and the gap between the waveguide and the disk is 300 nm. The waveguide is linearly tapered down to a width of 125 nm at its end for low loss coupling to/from optical fibers. A sacrificial silicon dioxide layer (≈ 600 nm) and a low stress silicon nitride layer (≈ 200 nm) are sequentially deposited in a low pressure chemical vapor deposition (LPCVD) furnace, patterned via optical lithography, and dry etched to form the nitride ring above the microdisk, nitride anchors and bridge to mechanically attach various structures and an electrical pad to ground the substrate. A photolithography step, buffered oxide etch (BOE), and ion implantation (Boron) process are used to dope the first part of the cantilever and the pads for electrical contacts to a doping level of $> 10^{19}$ cm$^{-3}$ for low contact resistivity. Another 1 μm oxide layer was deposited using a low temperature oxide (LTO) furnace. The wafer was annealed for 1 h at 1000 °C in an ambient $N_2$ environment. Another photolithography, metal (Ni) hard mask deposition and liftoff, RIE and TMAH wet etching process was used to define fiber V-grooves in the Si substrate. Silicon dioxide layers are removed by 49 % HF wet etching to undercut and release the movable structures. A critical point drying process is used to avoid stiction between parts due to capillary forces. In a self-aligned region at the end of each grove the on-chip waveguide inverse tapers are coupled to an optical fiber placed in the V-groove, actively aligned and cured into place with ultraviolet (UV) light curable epoxy. The fiber to fiber loss of the pigtailed device is 16 dB.

**Appendix B**

In this appendix, we describe the displacement and force spectra calculation process. The directly measured voltage signal is converted to the displacement noise by dividing it by the optomechanical gain. The optomechanical gain is a product of the optomechanical coupling $g_{om}$ and the slope of the cavity expressed as $dV_{det}/d\omega$, where $\omega$ is the optical frequency. $g_{om}$ is reported in Fig. 2d, while the slope is calibrated in two different ways. In a first method, we directly measure the detector voltage signal corresponding to a fixed known frequency modulation (rms amplitude: 7.8 MHz, at 1 kHz) of the laser wavelength. Note that if the damping feedback is engaged, the cantilever will respond to the varying detector voltage by moving, and so the closed-loop detector modulation will be larger, resulting from both the cantilever motion and the wavelength shift. The factor of modulation enhancement is directly related to the cantilever stiffness change and is equal to $(f/f')^2$, where $f$ and $f'$ are the open-loop and closed-loop resonance frequencies, respectively. The frequencies are obtained by fitting the measured transfer functions. In a second method, the optomechanical gain is directly calculated from the transfer function. A known low frequency (1kHz) voltage is applied to the actuator resulting in a known displacement based on Fig. 2a in the absence of feedback. Under closed-loop control the cantilever becomes effectively softer and the response at low frequency increases. The relative stiffness change is obtained by fitting the system transfer function to extract the closed-loop resonance frequency and then taking the ratio $(f'/f)^2$. The optomechanical gain is calculated by dividing the photodetector signal by the known cantilever displacement. Note, due to the small difference between DC and 1 kHz and the difficulty of measuring a signal at DC, we use the signal at 1 kHz to estimate that at DC. At all the voltage levels in Fig. 5, the differences between the two calibration methods are within 20%.

Transfer functions are calculated as a ratio of a given mechanical displacement to an applied force causing it. The displacement is obtained from the photodetector signal as described above. The equivalent force is obtained from the actuator AC voltage. An AC voltage at low frequency will move the cantilever by a known distance $Z$ based on data in Fig. 2a. The cantilever would also move by the same distance if a force $F=k_0Z$ were applied to the ring at the end of the cantilever. In the absence of feedback $k_0$ is taken to be 0.038 N/m based on FEM. The force equivalent to a given voltage is assumed independent of frequency based on the broadband nature of electrostatic actuation. This allows us to convert the measured ratios $V_{det}/V_{ac}(f)$ to the calibrated transfer functions.

The force noise spectra are obtained by dividing the calibrated displacement noise spectra by the calibrated transfer functions.

All the uncertainties quoted in the paper are one standard deviation unless specified otherwise. The quoted uncertainty for the displacement sensitivity is calculated based on the variation of the photo detector limited background over the frequency range and the variation between the two calibration methods. The force sensitivity uncertainty is extracted through statistical analysis of the photo detector limited force background in the frequency range of 24 kHz to 26 kHz. The uncertainty of the ratio of displacement sensitivity to SQL is dominated by the displacement sensitivity uncertainty.

**References**

[1] Su S X P, Yang H S and Agogino A M 2005 A resonant accelerometer with two-stage microleverage mechanisms fabricated by SOI-MEMS technology *IEEE Sensors Journal*. **5** 1214-23

[2] Krishnamoorthy U, Olsson III R H, Bogart G R, Baker M S, Carr D W, Swiler T P and Clews P J 2008 In-plane MEMS-based nano-g accelerometer with sub-wavelength optical resonant sensor *Sensors and Actuators A: Physical* **145-146** 283-90

[3] Sahin O, Magonov S, Su C, Quate C F and Solgaard O 2007 An atomic force microscope tip designed to measure time-varying nanomechanical forces *Nature Nanotech*. **2** 507-14

[4] Zotov S A, Rivers M C, Trusov A A and Shkel A M 2011 Folded MEMS pyramid inertial measurement unit *IEEE Sensors Journal*. **11** 2780-9.

[5] Arlett J L, Myers E B and Roukes M L 2011 Comparative advantages of mechanical biosensors *Nature Nanotech*. **6** 203-15

[6] Hammel P C and Pelekhov D V 2007 *The Magnetic Resonance Force Microscope. Handbook of Magnetism and Advanced Magnetic Materials* Volume 5: *Spintronics and Magnetoelectronics* (John Wiley & Sons, Ltd.)

[7] Giessibl F 2003 Advances in atomic force microscopy. *Rev. Mod. Phys*. **75** 949–83.

[8] Bolle C A, Aksyuk V, Pardo F, Gammel P L, Zeldov E, Bucher E, Boie R, Bishop D J and Nelson D R 1999 Observation of mesoscopic vortex physics using micromechanical oscillators *Nature* **399** 43-6

[9] Poggio M and Degen C L 2010 Force-detected nuclear magnetic resonance: recent advances and future challenges *Nanotech.* **21** 342001

[10] Zolfagharkhani G, Gaidarzhy A, Degiovanni P, Kettemann S, Fulde P and Mohanty P 2008 Nanomechanical detection of itinerant electron spin flip *Nature Nanotech.* **3** 720-3.

[11]   Mohideen U and Roy A 1998 Precision measurement of the Casimir force from 0.1 to 0.9 μm *Phys. Rev. Lett*. **81**, 4549–52

[12]   Chan H B, Aksyuk V, Kleiman R N, Bishop D J and Capasso F 2001 Quantum mechanical actuation of microelectromechanical systems by the Casimir force *Science* **291** 1941-4

[13]   Decca R S, Lopez D, Fischbach E, Klimchitskaya G L, Krause D E and Mostepanenko V M 2007 Tests of new physics from precise measurements of the Casimir pressure between two gold-coated plates *Phys. Rev. D* **75** 077101

[14]   Kippenberg T J and Vahala K J 2007 Cavity Opto-Mechanics *Opt. Express* **15** 17172-205

[15]   Kippenberg T J and Vahala K J 2008 Cavity Optomechanics: Back-Action at the Mesoscale *Science* **321** 1172-6

[16]   Hoogenboom B W, Frederix P L T, Yang J L, Martin S, Pellmont Y, Steinacher M, Zäch S, Langenbach E, Heimbeck H-J, Engel A and Hug H J 2005 A Fabry–Perot interferometer for micrometer-sized cantilevers *Appl. Phys. Lett*. **86** 074101

[17]   Abramovici A et al. 1992 LIGO: the laser interferometer gravitational-wave observatory *Science* **256** 325-33

[18]   Schliesser A, Rivie're R, Anetsberger G, Arcizet O and Kippenberg T J 2008 Resolved-sideband cooling of a micromechanical oscillator *Nature Phys*. **4** 415-19

[19]   Anetsberger G, Arcizet O, Unterreithmeier Q P, Rivière R, Schliesser A, Weig E M, Kotthaus J P, and Kippenberg K J 2009 Near-field cavity optomechanics with nanomechanical oscillators *Nature Phys*. **5** 909-14

[20]   Arcizet O, Cohadon P F, Briant T, Pinard M Heidmann A, Mackowski J M, Michel C, Pinard L, Français O and Rousseau L 2006 High-Sensitivity Optical Monitoring of a Micromechanical Resonator with a Quantum-Limited Optomechanical Sensor *Phy. Rev. Lett*. **97** 133601

[21]   Teufel J D, Donner T, Castellanos-Beltran M A, Harlow J W and Lehnert K W 2009 Nanomechanical motion measured with an imprecision below that at the standard quantum limit *Nature Nanotech*. **4** 820-3

[22]   Rakich P T, Popovic M A, Soljacic M and Ippen E P 2007 Trapping, corralling and spectral bonding of optical resonances through optically induced potentials *Nature Photon*. **1** 658-65

[23]   Favero I and Karrai K 2009 Optomechanics of deformable optical cavities *Nature Photon*. **3** 201-5

[24]   Chan J Mayer Alegre T P, Safavi-Naeini A H, Hill J T, Krause A, Gröblacher S, Aspelmeyer M and Painter O 2011 Laser cooling of a nanomechanical oscillator into its quantum ground state *Nature* **478** 89-92

[25]   Eichenfield M, Chan J, Camacho R M, Vahala K J and Painter O 2009 Optomechanical crystals *Nature* **462** 78-82

[26]   Bagheri M, Poot M, Li M, Pernice W P H and Tang H X 2011 Dynamic manipulation of nanomechanical resonators in the high-amplitude regime and non-volatile mechanical memory operation *Nature Nanotech*. **6** 726-32

[27]   Li M, Pernice W H P and Tang H X 2009 Reactive Cavity Optical Force on Microdisk-Coupled Nanomechanical Beam Waveguides *Phy. Rev. Lett.* **103** 223901

[28]   Wiederhecker G S, Chen L, Gondarenko A and Lipson M 2009 Controlling photonic structures using optical forces *Nature* **462** 633-6

[29]   Srinivasan K, Miao H, Rakher M T, Davanco M and Aksyuk V 2011 Optomechanical transduction of an integrated silicon cantilever probe using a microdisk resonator *Nano Lett*. **11** 791-7

[30]   Lee M M and Wu M C 2006 Tunable coupling regimes of silicon microdisk resonators using MEMS actuators *Opt. Express* **14** 4703-12

[31] Pruessner M W, Stievater T H and Rabinovich W S 2008 In-plane microelectromechanical resonator with integrated Fabry–Pérot cavity *Appl. Phys. Lett*. **92** 081101

[32] Frank I W, Deotare P B, McCutcheon M W and Lončar M 2010 Programmable photonic crystal nanobeam cavities *Opt. Express* **18** 8705-12

[33] Braginsky V B, Khalili F Y and Thorne K S 1995 *Quantum Measurement* (Cambridge University Press)

[34] Cohadon P F, Heidmann A and Pinard M 1999 Cooling of a Mirror by Radiation Pressure *Phys. Rev. Lett*. **83** 3174-7

[35] Poggio M, Degen C L, Mamin H J and Rugar D 2007 Feedback Cooling of a Cantilever's Fundamental Mode below 5 mK *Phys. Rev. Lett.* **99** 017201

[36] Lee K H, McRae T G, Harris G I, Knittel J and Bowen W P 2010 Cooling and Control of a Cavity Optoelectromechanical System *Phys. Rev. Lett*. **104** 123604

[37] Teufel J D, Donner T, Castellanos-Beltran M A, Harlow J W and Lehnert K W 2009 Nanomechanical motion measured with an imprecision below that at the standard quantum limit *Nat. Nano*. **4** 820-3.

[38] Genes C, Vitali D, Tombesi P, Gigan S and Aspelmeyer M 2008 Ground-state cooling of a micromechanical oscillator: Comparing cold damping and cavity-assisted cooling schemes *Phys. Rev. A* **77** 033804

[39] Borselli M, Johnson T and Painter O 2005 Beyond the Rayleigh scattering limit in high-Q silicon microdisks: theory and experiment *Opt. Express* **13** 1515-30

[40] Takahashi Y, Hagino H, Tanaka Y, Song B, Asano T and Noda S 2007 High-Q nanocavity with a 2-ns photon lifetime *Opt. Express* **15** 17206-13

[41] Schliesser A, Anetsberger G, Rivière R, Arcizet O and Kippenberg T J 2008 High-sensitivity monitoring of micromechanical vibration using optical whispering gallery mode resonators *New J. Phys*. **10** 095015

[42] Black E D 2001 An introduction to Pound–Drever–Hall laser frequency stabilization *Am. J. Phys.* **69** 79-87

[43] Yang J, Ono T and Esashi M 2000 Surface effects and high quality factors in ultrathin single-crystal silicon cantilevers *Appl. Phys. Lett.* **77** 3860

**Acknowledgements**

We thank the staff of the CNST Nanofab, especially Lei Chen and Richard Kasica, for assistance with fabrication and thank Alan Band for his assistance with circuits. HM acknowledges support under the Cooperative Research Agreement between the University of Maryland and the National Institute of Standards and Technology Center for Nanoscale Science and Technology, Award 70NANB10H193, through the University of Maryland. This project was supported in part by the DARPA MESO program.

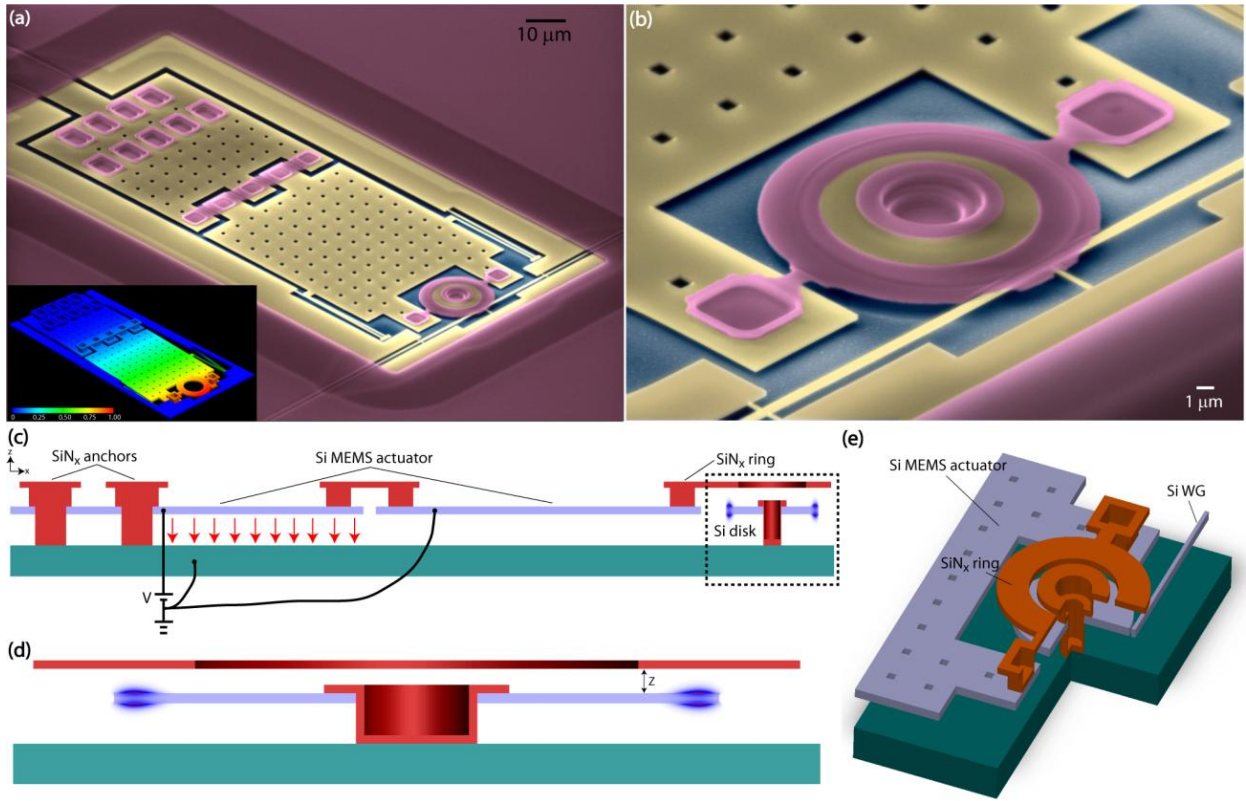

Figure 1 Device geometry. (a) Scanning electron micrograph (SEM) of the fabricated device with color indicating different material layers. (inset) Finite element simulation of the fundamental mechanical mode of the mechanical structure (actuator plus silicon nitride ($SiN_x$) ring). (b) Zoomed-in view of the sensor area. (c)-(d) 2D and (e) 3D cross-sectional illustrations of the device. The illustration shown in (d) is a zoomed-in cross section of the microdisk resonator and $SiN_x$ ring. The illustration shown in (c) is a cross section taken through the key elements of the mechanical transducer and optical sensor (i.e., $SiN_x$ anchors, silicon (Si) MEMS actuator, $SiN_x$ ring, and Si disk). In these cross-section illustrations, the $Z$ axis scale has been adjusted for clarity.

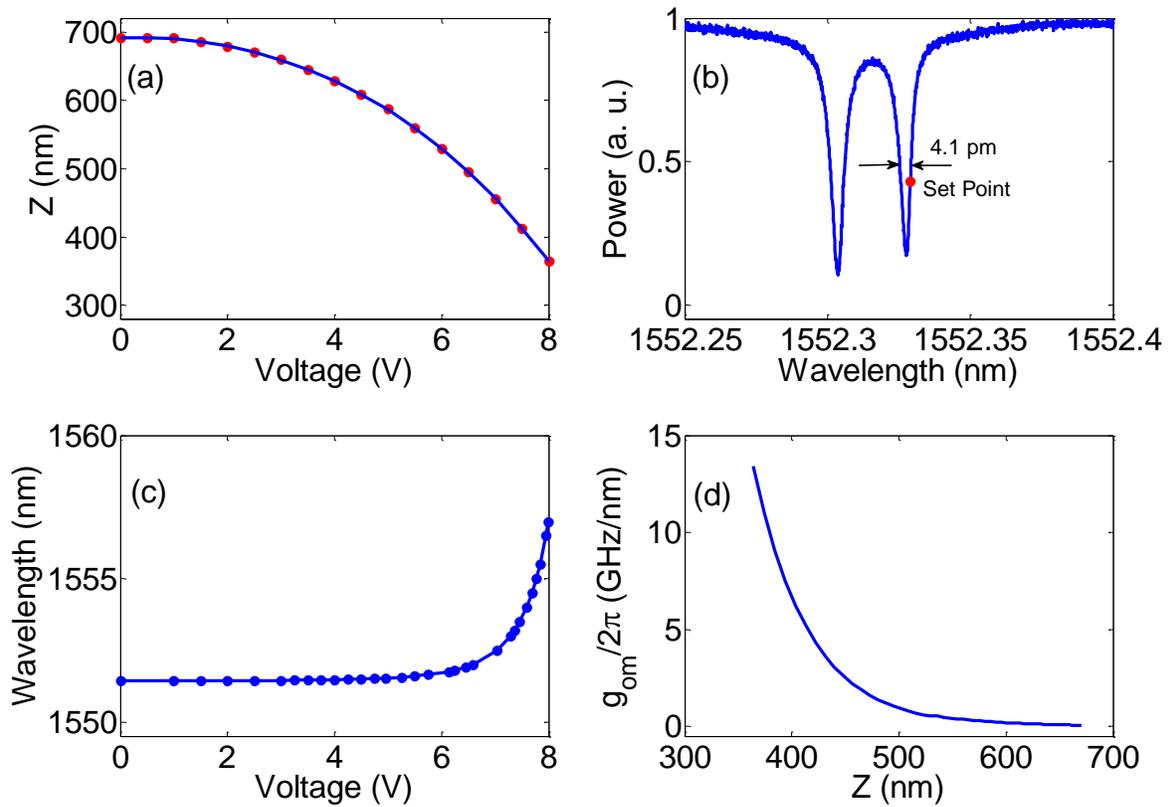

Figure 2 Static properties of the system. (a) Displacement as a function of bias voltage. (b) Transmission spectrum showing the optical mode. (c) Resonant wavelength tuning as a function of bias voltage. (d) Optomechanical coupling as a function of displacement. In (a) and (c) the dots are the positions where the measurements were made.

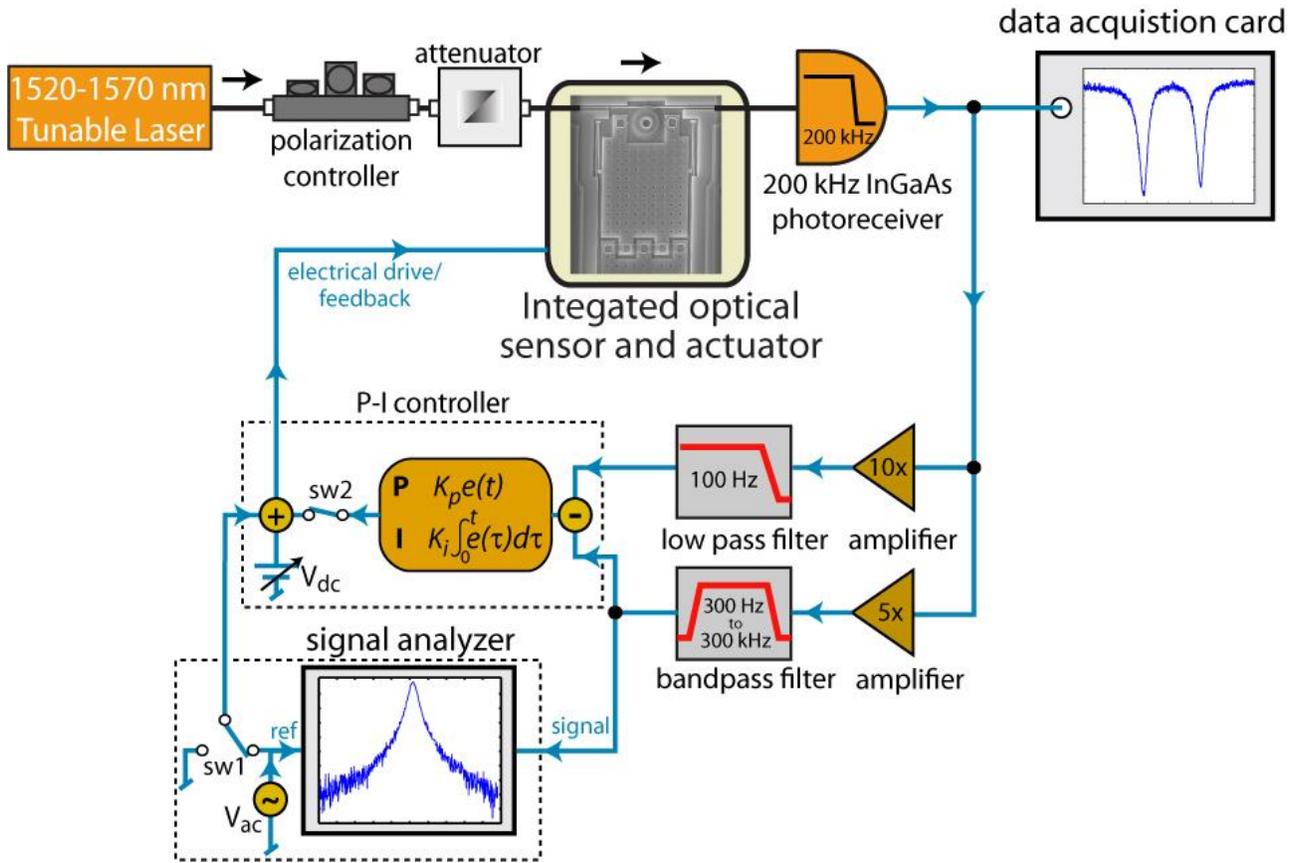

Figure 3 Schematic illustration of the experimental setup and the feedback control system. The displacement spectra and the transfer functions are measured by disconnecting and connecting the AC supply (indicated with the switch sw1), with feedback on or off (switch sw2). The top, low frequency feedback arm is for cavity locking, the bottom feedback arm is for cold-damping.

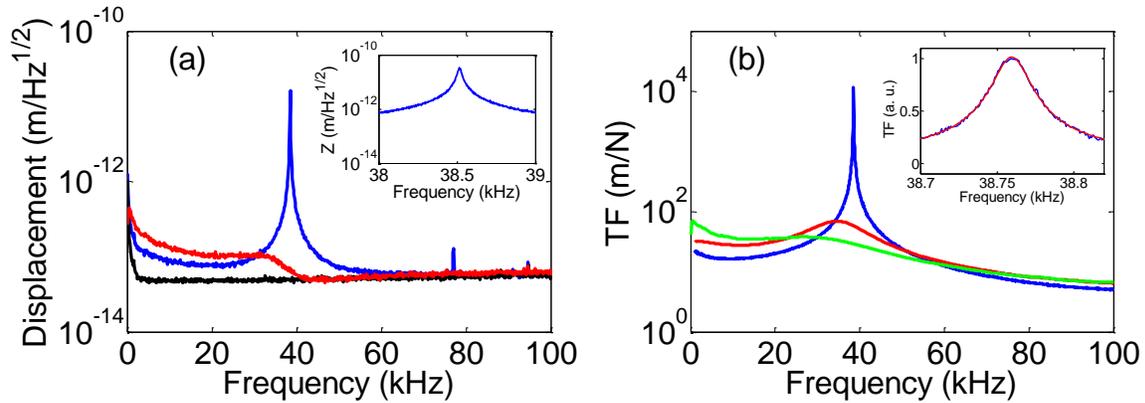

Figure 4 Device dynamics with and without cold damping. (a) Displacement noise spectra at 3 V DC bias with (red) and without (blue) feedback cooling. (b) Transfer functions (TF) corresponding to (a) (blue and red), and with higher feedback gain (green). Black curve in (a) shows the photodetector limited measurement sensitivity for both the undamped case and the damped case which are not distinguishable. The small second harmonic peak at ≈ 77 kHz (blue curve in a) indicates that at this bias level, without feedback cooling, the displacement noise already slightly exceeds the sensor linear dynamic range. The spectrum in (a) is measured with a resolution bandwidth of 128 Hz, wider than the mechanical resonance without feedback, therefore the points near resonance do not reflect the true values.. The inset in (a) shows the displacement spectrum near resonance measured with a resolution bandwidth of 2 Hz. The inset in (b) shows the transfer function near resonance measured at low optical power (blue) and a fit (red) corresponding to the mechanical $Q$ of 1370.

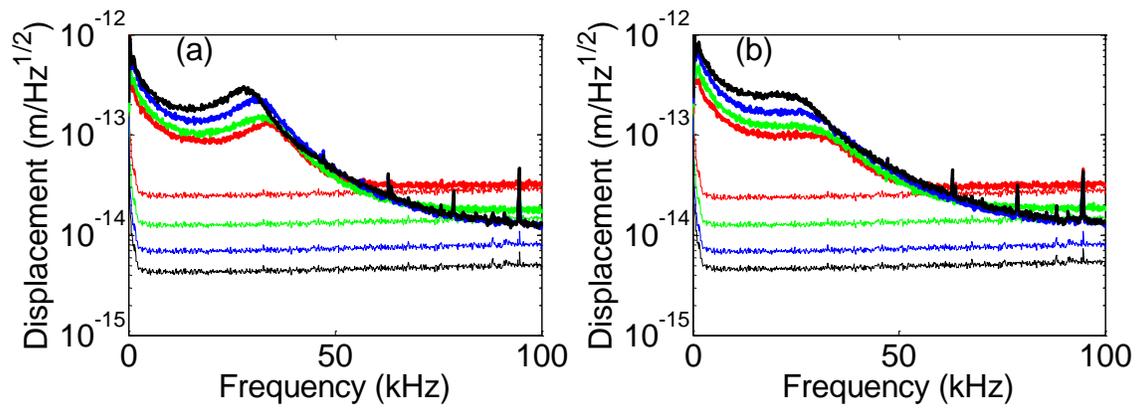

Figure 5 Displacement spectra at various bias voltages. (a) Under-damped case. (b) Critically damped case. The bias voltages for the red, green, blue and black curves are 4.5 V, 5.5 V, 6.25 V and 6.75 V, respectively. The thicker lines show the measured ring displacement noise, the thinner lines are the corresponding photodetector limited background curves. The highest achieved readout sensitivity corresponds to the thin black lines.

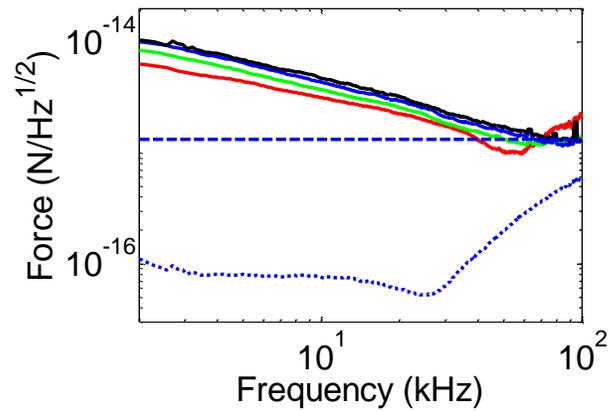

Figure 6 Force noise spectra at various bias voltages corresponding to Fig. 5b. The bias voltages for the red, green, blue and black curves are 4.5 V, 5.5 V, 6.25 V and 6.75 V, respectively. The dashed horizontal line is the estimated thermal force, while the dotted line is the photo detector limited force measurement background at 6.75 V bias voltage. Highest force sensitivity corresponds to the dashed line values around 25 kHz.